# Spatially-correlated Site Occupancy in the Nonstoichiometric Meta-stable $\varepsilon$-Al$_{60}$Sm$_{11}$ Phase during Devitrification of Al-10.2 at.% Sm Glasses

Lin Yang[1,2], Feng Zhang[2,a)], Fan-Qiang Meng[2], Lin Zhou[2], Yang Sun[2], Xin Zhao[2], Zhuo Ye[2], Matthew J. Kramer[2], Cai-Zhuang Wang[2], Kai-Ming Ho[1,2,b)]

[1]*Department of Physics, Iowa State University, Ames, Iowa, 50011, USA*

[2]*Ames Laboratory, Ames, Iowa, 50011, USA*

a) Corresponding author: fzhang@amesla.gov

b) Corresponding author: kmh@iastate.edu

**Keyword:** *nonstoichiometric compound; cluster expansion; Monte Carlo simulation; molecular dynamics simulation*

**Abstract**

A metastable $\varepsilon$-Al$_{60}$Sm$_{11}$ phase appears during the initial devitrification of as-quenched Al-10.2 at.% Sm glasses. The $\varepsilon$ phase is nonstoichiometric in nature since Al occupation is observed on the 16$f$ Sm lattice sites. Scanning transmission electron microscopic images reveal profound spatial correlation of Sm content on these sites, which cannot be explained by the "average crystal" description from Rietveld analysis of diffraction data. Thermodynamically favorable configurations, established by Monte Carlo (MC) simulations based on a cluster-expansion model, also give qualitatively different correlation functions from experimental observations. On the other hand, molecular dynamics simulations of the growth of $\varepsilon$-Al$_{60}$Sm$_{11}$ in undercooled liquid show that when the diffusion range of Sm is limited to ~ 4 Å, the correlation function of the as-grown crystal structure agrees well with that of the STEM images. Our results show that kinetic effects, especially the limited diffusivity of Sm atoms plays the fundamental role in determining the nonstoichiometric site occupancies of the $\varepsilon$-Al$_{60}$Sm$_{11}$ phase during the crystallization process.

## 1. Introduction

Al alloyed with ~ 10 at.% Sm represents a typical Al-rare earth (RE) system that can undergo deep undercooling from liquid and form amorphous solids or nanocrystalline composite materials with much improved mechanical properties compared with pure Al [1–4]. When the as-quenched amorphous Al-10.2 at.% Sm melt-spun ribbons are heated, a meta-stable cubic phase is usually the first phase to appear in the multiple-step devitrification process [5]. Although this cubic phase was reported more than 20 years ago [6,7], not until recently has its atomic structure been solved by an approach integrating experimental diffraction data, a genetic algorithm for crystal structure prediction, and molecular dynamics (MD) simulations [5]. The solved phase, labelled as $\varepsilon$-Al$_{60}$Sm$_{11}$, has a body-centered cubic (BCC) unit cell (space group $Im\bar{3}m$, No. 229) with a lattice constant of 13.9 Å. For the stoichiometric Al$_{60}$Sm$_{11}$ crystal, each cubic unit cell contains 120 Al atoms and 22 Sm atoms. However, both Rietveld analysis and MD simulations show that, among the 22 Sm sites, only 6 sites are fully occupied by Sm, and the remaining sites with a Wyckoff notation 16$f$ are shared with Al atoms, resulting in a non-stoichiometric phase Al$_{60+x}$Sm$_{11-x}$ with $x$ ~ 4 (see Fig. 1a).

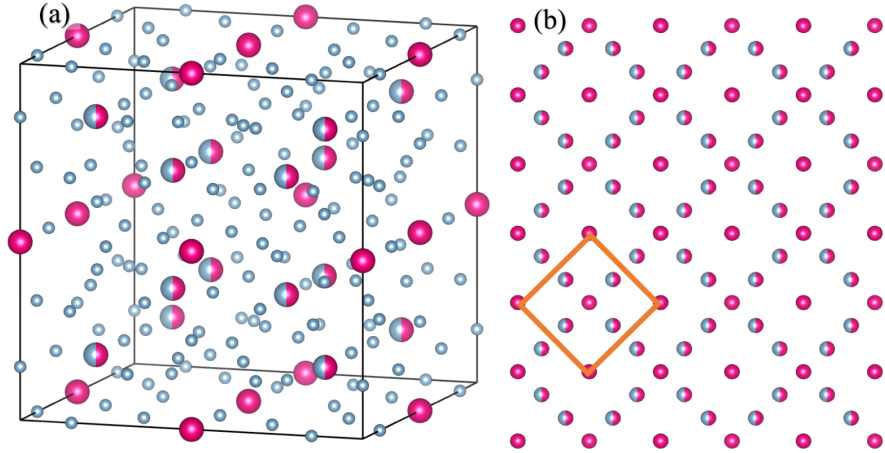

**Fig. 1.** (a) The unit cell of the $\varepsilon$-$Al_{60}Sm_{11}$ phase. Red and blue denote Sm and Al atoms, respectively. The half red and half blue sites can be occupied by either Al or Sm (PO Sm sites). (b) Top view of the FO and PO Sm sites of the [001] plane. The orange square shows a unit cell of the projected 2D lattice.

In Ref. [5], the authors argue that the tolerance to Al occupation on the PO sites is an important reason that the $\varepsilon$ phase can readily nucleate and grow. However, it remains obscure what is the underlying reason for the shared occupation. Traditionally, nonstoichiometry in inorganic compounds is understood from thermodynamic factors such as enthalpy of formation and configurational entropy [8], while the kinetics associated with processing conditions is often overlooked. In this paper, we collect detailed information about how the Sm/Al atoms occupy the PO sites in the $\varepsilon$ phase from scanning transmission electronic microscopy (STEM), which clearly shows that the occupancies of the PO sites are spatially correlated across multiple unit cells. Monte Carlo (MC) and molecular dynamics (MD) simulations show that the spatial correlation is not originated from the thermodynamically favorable configurations. Instead, kinetic effects, especially the limited diffusivity of Sm atoms plays an important role in creating the observed spatial correlations.

## 2. Experiments

To further characterize the morphology of the $\varepsilon$ phase, the specimens are quenched to room temperature, then they are prepared for scanning transmission electron microscopy (STEM) characterization by focused ion beam (FIB) milling until the specimen thickness is as thin as around 50~100 nm. The FIB instrument in use is the FEI dual-beam Helios NanoLab G3 UC, and a FEI Titan Themis 300 Cube aberration corrected scanning transmission electron microscope. The high angle annular dark-field scanning transmission electronic microscopic (HAADF-STEM) is used to characterize the atoms projected onto the [001] crystal plane of the specimens, which has a square lattice with $a = 9.83$ Å, as shown in orange box in Fig. 1b.

Fig. 2a shows a typical HAADF-STEM image, where only Sm atoms are clearly seen. This is reasonable since Sm has a much higher electron density than Al. Fig. 2b zooms in a portion of Fig. 2a, in which 2 PO sites (PO1 and PO2) and 1 FO site within a unit cell are labelled along the <110> direction. The intensity at each Sm site is generally proportional to the total number of Sm atoms in the projected column. However, the raw image of Fig. 2a cannot be directly used to compare Sm occupancies at different sites, because of the non-uniform background intensities, which is indicative of fluctuations of thickness and composition in the sample. On the other hand, it is reasonable to assume that such fluctuation is negligible within one unit cell. Thus, one can use the intensity at the FO site to calibrate the intensity at other sites of the same unit cell along a <110> line. More specifically, we define a normalized intensity as $I_n = I/(2I_{FO})$, where $I$ is the absolute intensity of an atomic site directly read from the raw image such as Fig. 2a, and $I_{FO}$ is the absolute intensity of the FO site in the same unit cell. A factor of 2 is added in the denominator because the number of PO sites doubles that of the FO sites in a projected column. Fig. 2c shows the normalized intensity along <110> lines of the same region as that of Fig. 2b.

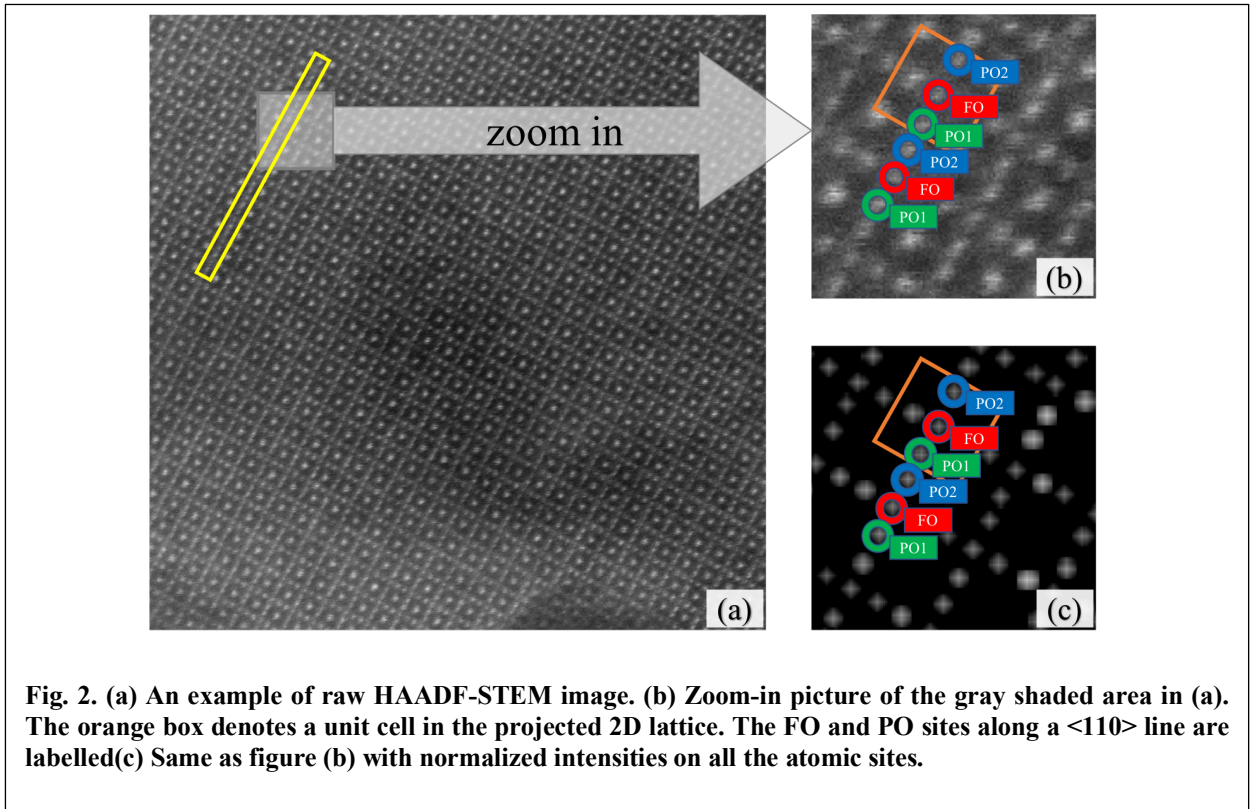

**Fig. 2. (a) An example of raw HAADF-STEM image. (b) Zoom-in picture of the gray shaded area in (a). The orange box denotes a unit cell in the projected 2D lattice. The FO and PO sites along a <110> line are labelled(c) Same as figure (b) with normalized intensities on all the atomic sites.**

Fig. 3a shows the normalized intensity along a <110> crystal line (circled by a yellow box in Fig. 2a), which indicates a large variation of the occupancy on the PO sites. Interestingly, if the difference of $I_n$ between two neighboring PO sites, PO1 and PO2 as shown in Fig. 2c, $I_d = I_{n,PO1} - I_{n,PO2}$ is plotted versus the unit cell number along the same line (Fig. 3b), one can see an oscillatory behavior with a period of multiple unit cells, suggesting some spatial correlations in the PO sites occupancy.

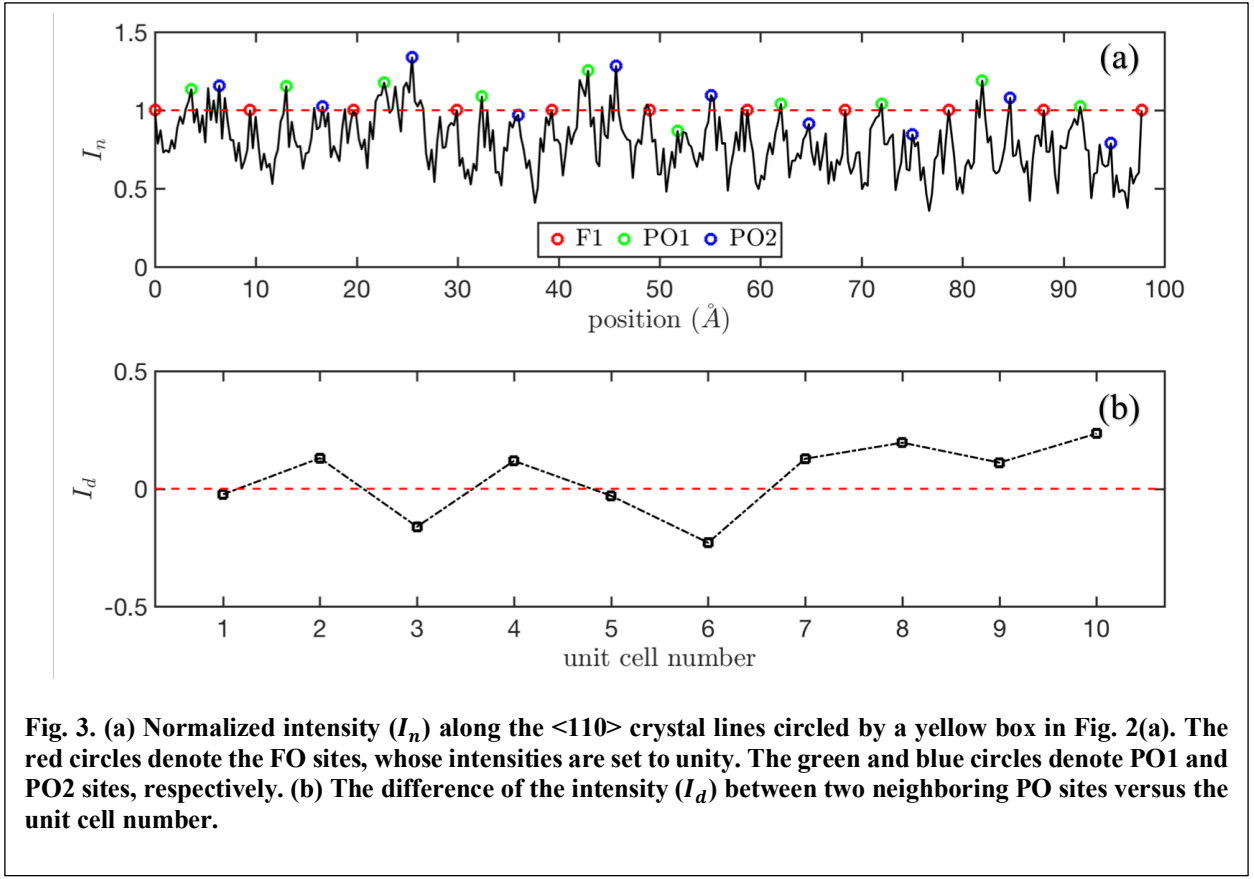

**Fig. 3. (a)** Normalized intensity ($I_n$) along the <110> crystal lines circled by a yellow box in Fig. 2(a). The red circles denote the FO sites, whose intensities are set to unity. The green and blue circles denote PO1 and PO2 sites, respectively. **(b)** The difference of the intensity ($I_d$) between two neighboring PO sites versus the unit cell number.

To better capture the spatial correlation, we define the correlation function between two unit cells along the <110> direction as:

$$g(\mathbf{R}) = \langle I_d(\mathbf{r}) I_d(\mathbf{r} + \mathbf{R}) \rangle_{\mathbf{r}}, \quad (1)$$

where **r** is the position of a PO site, **R** is a lattice vector along the <110> direction in the projected 2D lattice (see Fig. 2b), and $\langle \cdots \rangle_{\mathbf{r}}$ denotes the average over all different positions **r**. In Fig. 4, we show the correlation function $g(R)$ averaged over all <110> lines in 6 STEM images collected from 2 Al-~10 at.% Sm specimens. For comparison purpose, we also show the correlation function generated by assuming the occupancy of Sm atoms on the PO sites is completely random. The solid lines are fittings to the decay function $g(R) = A e^{-R/\xi}$. According to Eq. (1), $A = g(0) = \langle I_d^2(\mathbf{r}) \rangle$, which essentially measures the variation ($\langle I_d^2 \rangle - \langle I_d \rangle^2$) of the distribution of Sm occupancy in all the PO sites since $\langle I_d \rangle^2$ is very small (<10$^{-4}$) in all the samples studied in this paper. $\xi$ is a parameter characterizing the correlation length. The value of $A$ is $0.020 \pm 0.002$ and $0.0062 \pm 0.0002$ for the STEM and random samples, respectively, indicating that the variation of Sm occupancy in experimental samples is significantly larger than that from a random sample. The correlation length $\xi$ for the experimental sample is $0.98\,a \pm 0.25\,a$, demonstrating non-vanishing correlations over multiple lattice constants. In contrast, $\xi$ for the random sample is merely $0.25\,a \pm 0.11\,a$, showing that the correlation is essentially zero since $R/a$ is a positive interger number.

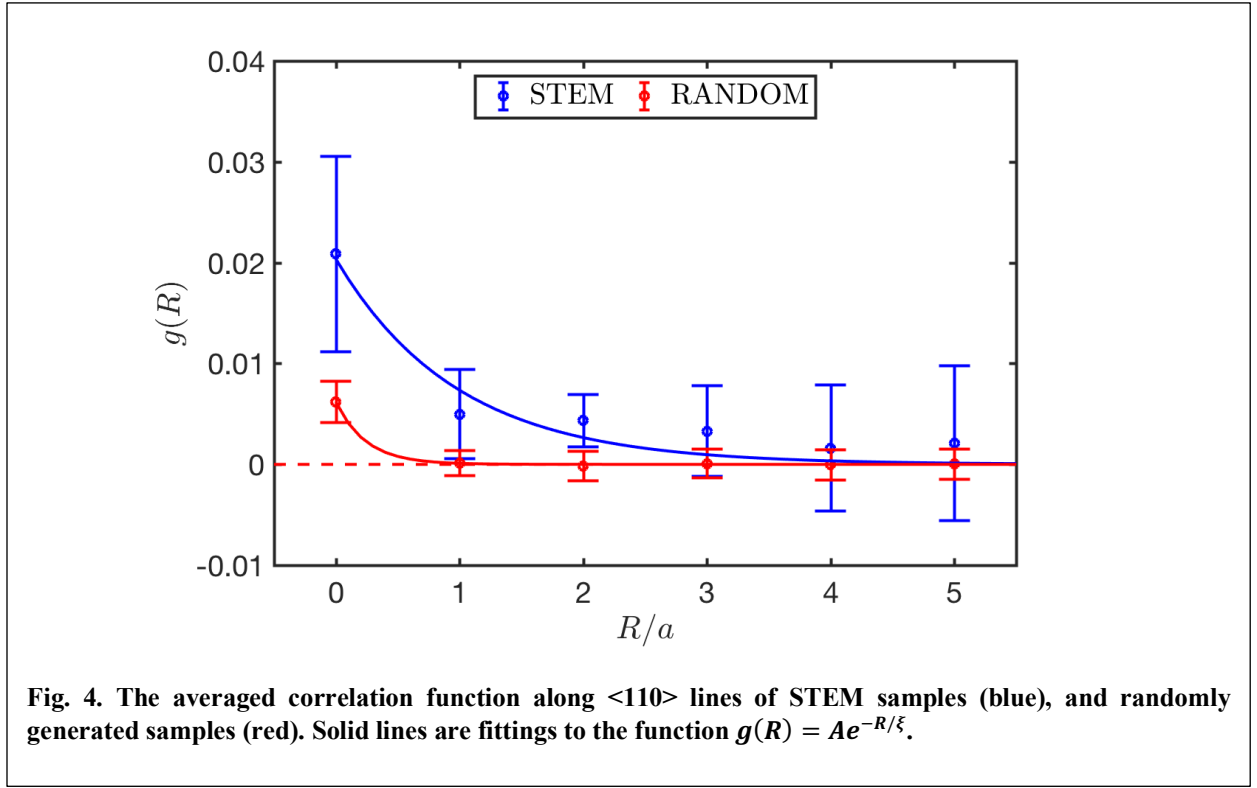

**Fig. 4.** The averaged correlation function along <110> lines of STEM samples (blue), and randomly generated samples (red). Solid lines are fittings to the function $g(R) = Ae^{-R/\xi}$.

To understand the origin of the correlation of site occupancy observed in experiments, we perform Monte Carlo (MC) simulation [9,10] based on a cluster expansion (CE) model [11,12], as well as classical molecular dynamics (MD) simulation [13,14]. MC simulations emphasize on establishing the thermodynamically favorable configurations; while kinetics especially the effects Sm mobility is addressed in MD simulations.

## 3. MC simulations

Since there is a well-defined underlying BCC lattice, and the partial occupancy is only associated with certain fixed lattice sites (i.e., the 16*f* PO sites), the free energy of the system can be regarded as a function determined only by the configurations of different chemical elements on the PO sites. Cluster expansion (CE) is an appropriate method to establish the free energy model for this system [11]:

$$F = F_0 + \sum_i J_i S_i + \sum_{i<j} J_{ij} S_i S_j + \sum_{i<j<k} J_{ijk} S_i S_j S_k + \cdots \quad (2)$$

where $S_i$ is the chemical occupation of the partially ocupied sites, which is set to be 1 (-1) if the site $i$ is occupied by Al (Sm), $J_i, J_{ij}, J_{ijk}, \ldots$ are the cluster expansion (CE) coefficients, and $F$ is the total free energy of the system which includes the internal energy and a contribution from the vibrational entropy. $F$ in Eq. (2) is expanded as a summation of contributions from all "clusters" including singlets, doubles, triplets, etc. In practice, only the first few terms are included in Eq. (2) assuming that higher order clusters have negligible contributions to $F$.

To fit the cluster expansion coefficients, we first calculate the free energy of 18 training structures with various occupation of the PO sites. The free energies of the training structures are calculated within the quasi-harmonic approximation [15,16] at two different temperatures $T$ = 500 K and 800 K. To further expedite the calculations, we have employed a classical potential in the Finnis-Sinclair (FS) form [17], which was carefully fitted to the ab-initio energetics of a series of Al-Sm compounds with ~ 10 at% Sm as well as the pair correlation functions of $Al_{90}Sm_{10}$ liquid [18]. Using this potential, we were able to identify the $\varepsilon$-$Al_{60}Sm_{11}$ phase in a genetic algorithm search. Moreover, the X-ray diffraction pattern of the $\varepsilon$ phase with built-in antisite defects directly grown in molecular dynamics simulations using the same potential matches excellently with experiments, suggesting that this interatomic potential also correctly captures the nature of defects in $\varepsilon$-$Al_{60}Sm_{11}$ [5]. With these data in hand, we then use the Alloy Theoretic Automated Toolkit (ATAT software) [19] to fit the CE coefficients.

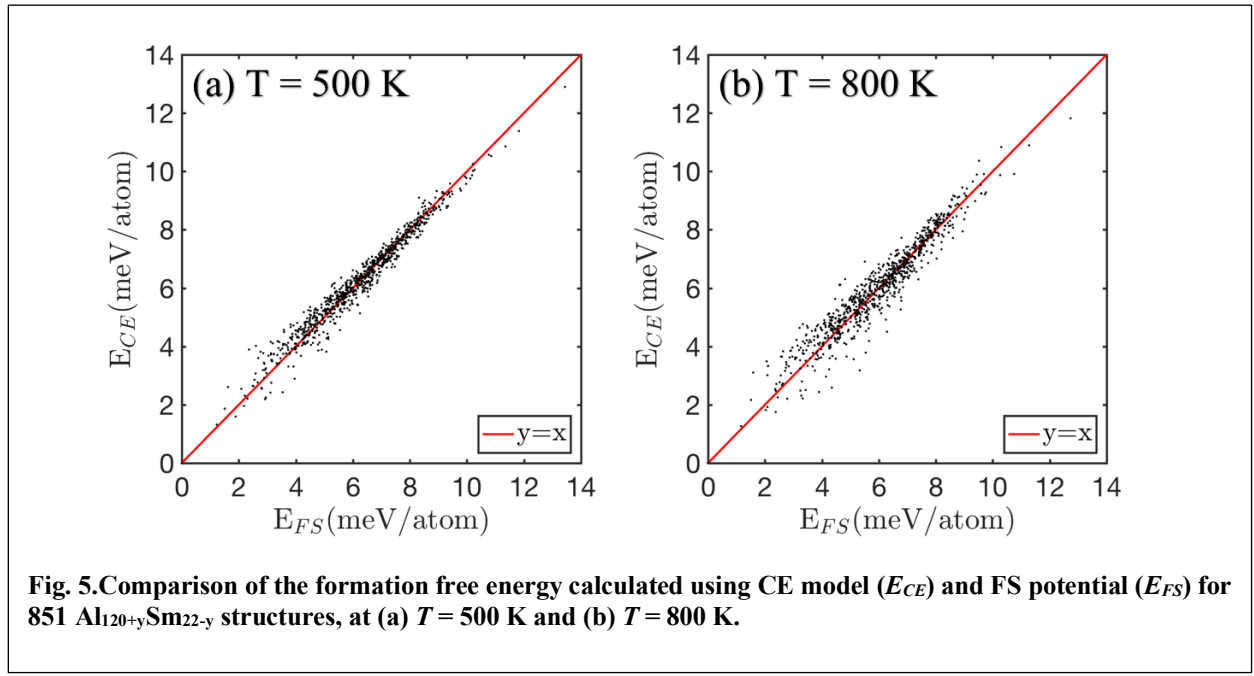

**Fig. 5.** Comparison of the formation free energy calculated using CE model ($E_{CE}$) and FS potential ($E_{FS}$) for 851 $Al_{120+y}Sm_{22-y}$ structures, at (a) $T$ = 500 K and (b) $T$ = 800 K.

As a validation of the CE model, we apply our model to compute the formation free energy of the structures of a unit cell ($Al_{120+y}Sm_{22-y}$) with different compositions and PO sites configurations, using the two end compositions $y = 0$ and $y = 16$ as references. These values are compared with those directly calculated by quasi-harmonic approximation based on the FS potential. A total of 851 inequivalent structures containing 142 atoms in a cubic unit cell are used in this step. The comparison is shown in Fig. 5a and Fig. 5b, for $T$ = 500 K and 800 K, respectively. We define the root mean square discrepancy (RMSD) to quantitatively evaluate the error of the CE model:

$$\text{RMSD} = \sqrt{\frac{1}{N}\sum_{i=1}^{N}\left(E_{CE}^i - E_{FS}^i\right)^2} \quad (3)$$

where $N$ is the total number of the structures (in this case, $N = 851$), $E_{CE}^i$ and $E_{FS}^i$ are the formation free energy of structure $i$ computed using the CE model and directly from the FS potential, respectively. RMSD = 0.34 (0.48) $meV/atom$ for $T = 500$ (800) K. Both values are several orders of magnitude smaller than the thermal fluctuation $k_B T$ at corresponding temperatures, indicating that the error of the CE model is well controlled.

Based on the validated CE model, we simulate the equilibrium configuration of the PO sites of the $\varepsilon$-$Al_{60}Sm_{11}$ phase at 500K and 800K, using the Metropolis MC algorithm [9]. 500 K is close to the temperature (468 K) at which the spatial correlation is observed in experiments, and 800 K is used for comparison with MD simulations that will be shown later. Since we only need to explicitly consider the 16 PO sites in a unit cell with size ~ 1.4 nm, we can create a fairly large simulation box containing $20 \times 20 \times 20 = 8,000$ unit cells (128,000 PO sites), with a box size of 27.8 nm. Initially, equal number of Sm and Al atoms are randomly distributed in the PO sites, corresponding to an overall composition of $Al_{64}Sm_7$, which is close to that of the experiments. Fig. 6a and 6b show the free energy as a function of Monte Carlo (MC) step at 500 K and 800 K, respectively. Both figures level off at the end of the simulation, indicating that simulations have reached the thermodynamic equilibrium.

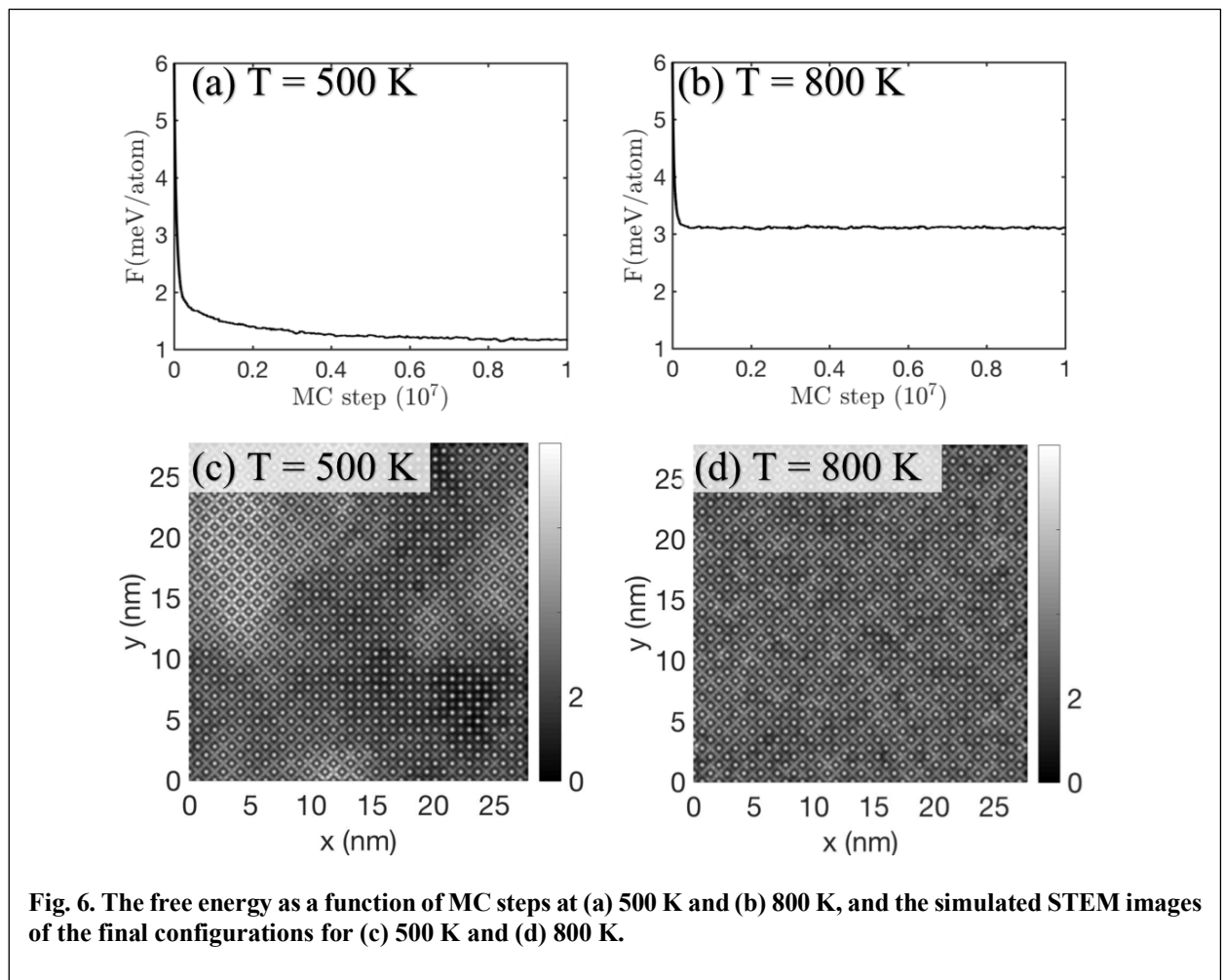

**Fig. 6. The free energy as a function of MC steps at (a) 500 K and (b) 800 K, and the simulated STEM images of the final configurations for (c) 500 K and (d) 800 K.**

To generate a "simulated" STEM image, we project the MC samples along the <001> direction, and make the intensity at each PO site directly proportional to the number of Sm atoms in the projected column (see Fig. 6c and 6d). Then, the <110> lines in the projected image are collected to analyze the correlation of Sm occupancies in the same way as we did on the experimental samples. However, the samples in our MC simulations are still significantly smaller than the experimental samples: the thickness of the MC samples along the projection direction is 27.8 nm, while that of experimental samples ranges from 50 to 100 nm. Here, a bootstrapping technique [20] is used to better match the experimental parameters. That is, two <110> lines randomly collected from the MC sample are stacked together to reproduce a "bootstrapped" sample with a thickness of 55.6 nm that is directly comparable with experiments. The number of total <110> lines that can be used for statistical analysis is also increased to $\binom{N}{2}$, where $N$ is the number of lines originally collected from the MC sample.

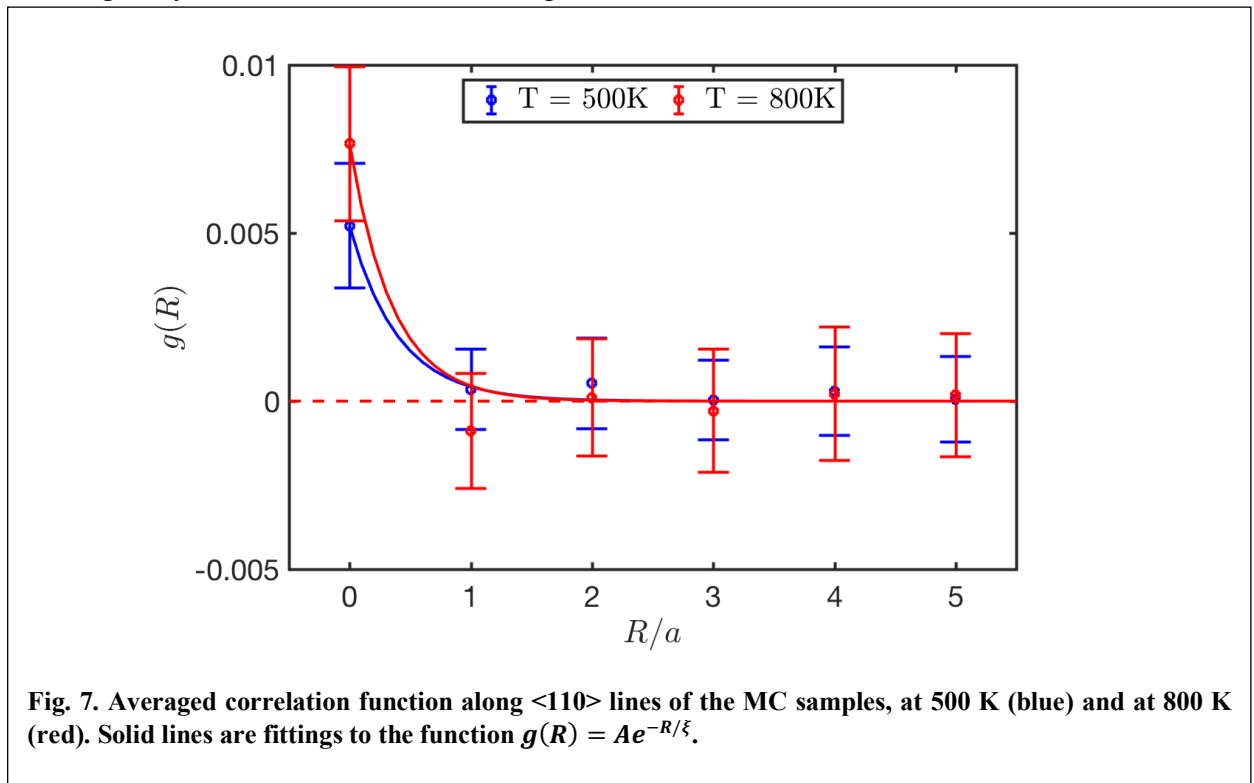

**Fig. 7. Averaged correlation function along <110> lines of the MC samples, at 500 K (blue) and at 800 K (red). Solid lines are fittings to the function $g(R) = Ae^{-R/\xi}$.**

The correlation functions $g(R)$ as defined in Eq. (1) of the bootstrapped MC samples at both 500 K and 800 K are given in Fig. 7. Again, the decay function $Ae^{-R/\xi}$ is used to fit the data. The fitting parameters at both temperatures are much closer to the randomly generated sample than to the STEM samples (see Fig. 4). This clearly demonstrates that the thermodynamically equilibrated configuration does not reproduce the observed spatial correlation in the experiments.

## 4. Molecular Dynamics Simulation

We then run MD simulations using the GPU-accelerated LAMMPS code [21,22] to simulate the growth of the ε-phase. The same classical potential as the one used to construct the cluster expansion model in MC simulations is used in the simulations [18]. Unlike MC, MD directly integrates Newtonic equations of motion and automatically include the dynamical effects. On the other hand, unlike our MC simulations which only consider the atoms occupying the PO sites, all atoms must be explicitly treated in MD simulations. Therefore, we can only afford to simulate a much smaller system. In our simulations, the initial configuration is constructed by inserting a seed layer of the $\varepsilon$ phase into the undercooled liquid (see Fig. 8a), which has a Sm concentration of 10.2

% and is pre-equilibrated. Periodic boundary conditions are also used in the initial configuration. The seed layer contains $4 \times 4 \times 1$ unit cells, in which the PO sites are randomly occupied by equal number of Sm and Al atoms. Since at the experiment temperature of ~ 500 K, the growth kinetics is too slow to be accessed within MD time scales, we run MD simulations at an elevated temperature of 800 K. At the same time, to better account for the limited Sm diffusivity at lower temperatures, we also run the "constrained" MD simulations, in which the diffusion of a Sm atom is limited to within a sphere of radius $r_c$ from its initial position in the liquid phase, by applying an additional potential $V(\mathbf{r}) = k|\mathbf{r} - \mathbf{r_0}|^6$. $\mathbf{r}$ is the position of a Sm atom and $\mathbf{r_0}$ is its initial position in the liquid phase. A large exponent 6 is used to ensure a flat bottom of $V(\mathbf{r})$ when $\mathbf{r}$ is close to $\mathbf{r_0}$. $k$ is varied to give two different values of $r_c$: 4.2 Å and 2.6 Å for the current work. These results are compared with those from unconstrained MD simulations ($r_c = \infty$). The final atomic structures for the MD simulations are shown in Fig. 8b-d. In all the cases, most of the liquid region has been transformed into the $\varepsilon$ phase. The growth essentially stops when the two growth fronts start to merge due to the periodic boundary conditions.

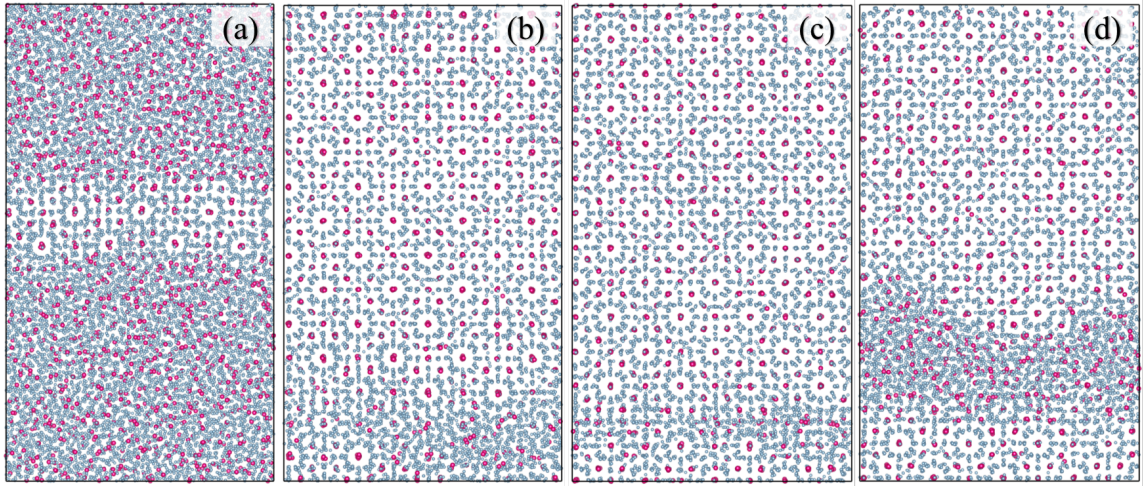

**Fig. 8. Atomic configurations during molecular dynamic (MD) simulations: (a) initial configuration; (b) final configuration of unconstrained MD ($r_c = \infty$); (c) final configuration for $r_c = 4.2$ Å; (d) final configuration for $r_c = 2.6$ Å.**

The thickness of the MD samples along the <001> direction is 5.56 nm. Again, the bootstrapping technique is used to generate a fair comparison with experiments. The correlation functions $g(R)$ of the three MD samples with different $r_c$ are shown in Fig. 9.

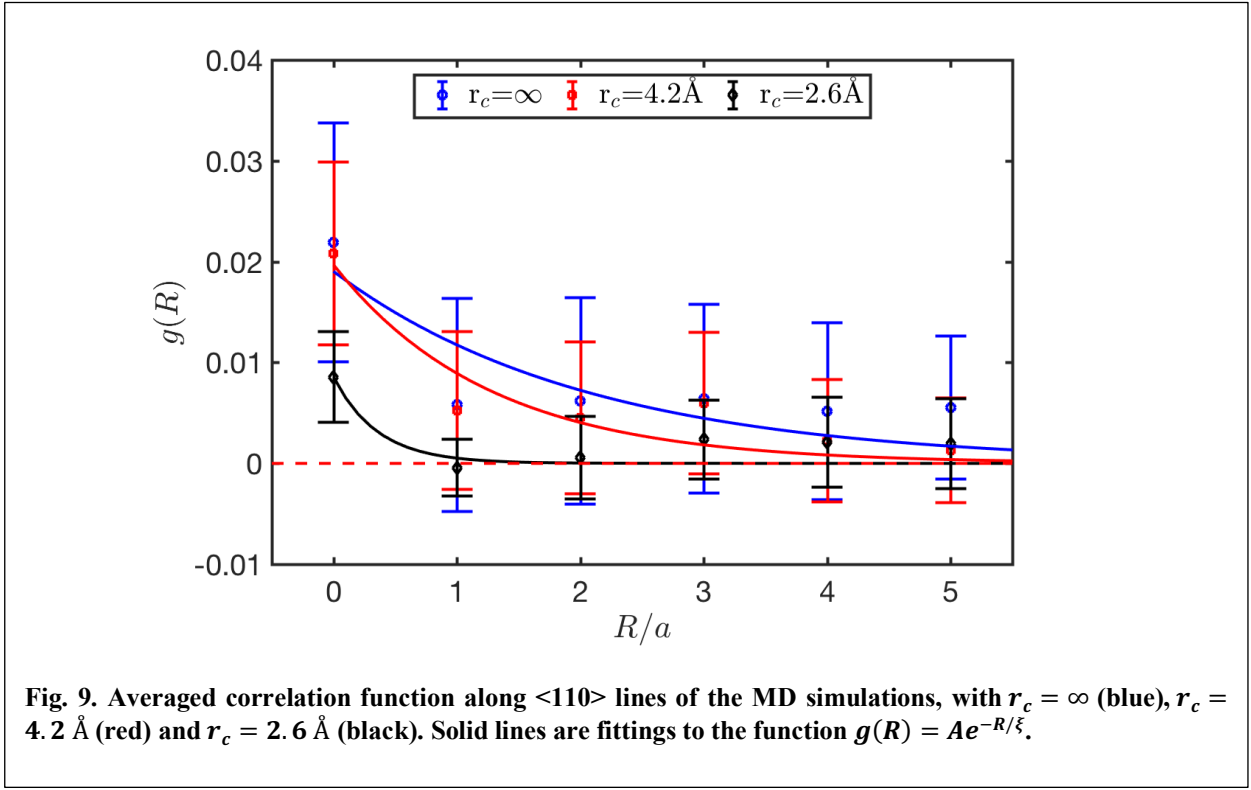

Fig. 9. Averaged correlation function along <110> lines of the MD simulations, with $r_c = \infty$ (blue), $r_c = 4.2$ Å (red) and $r_c = 2.6$ Å (black). Solid lines are fittings to the function $g(R) = Ae^{-R/\xi}$.

In Fig. 10, we show the estimated fitting parameters $A$ and $\xi$, together with their standard errors for all the samples studied in this paper. One can see that only the sample generated in MD simulations with $r_c = 4.2$ Å has a significant overlap with the STEM samples, indicating that limited Sm mobility plays an important role in producing the special correlation in the PO site occupancy. However, when the Sm movement is over confined ($r_c = 2.6$ Å), the Sm distribution on the PO sites carries too much legacy from the random liquid structure, which is characterized by much smaller variance and essentially no inter-unit cell correlation. On the other hand, the MD sample with $r_c = \infty$ clearly overestimates the correlation length. It should be noted that even with $r_c = \infty$, the diffusion of Sm atoms in MD simulations is still local in nature due to its intrinsic low diffusivity. This is different from the MC simulations at the same temperature $T = 800$ K, in which long range diffusion of Sm atoms is needed in order to establish the thermodynamic equilibrium. As a result, one can see strong contrast of the fitting parameters between the MC sample at 800 K and the MD sample with $r_c = \infty$.

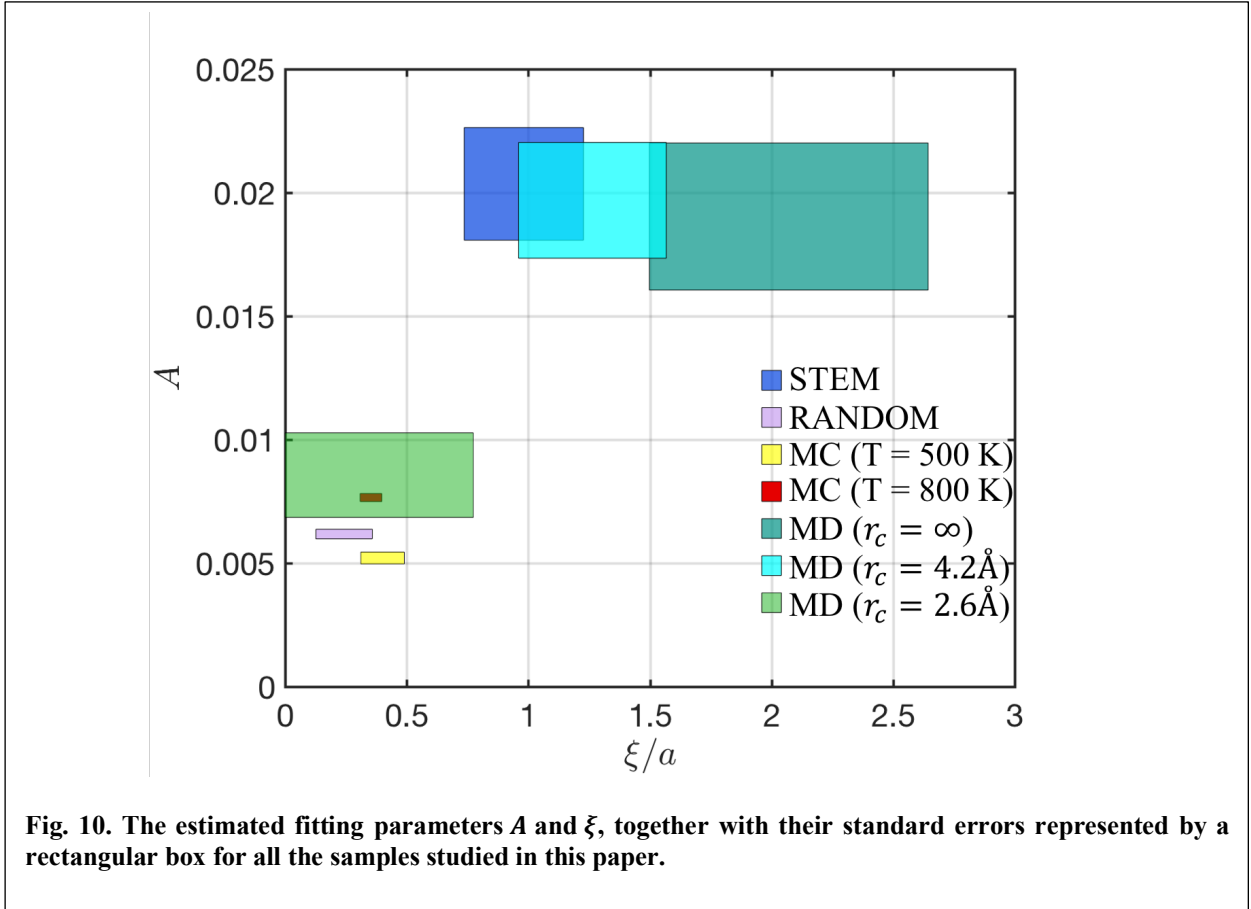

**Fig. 10.** The estimated fitting parameters *A* and *ξ*, together with their standard errors represented by a rectangular box for all the samples studied in this paper.

## 5. Conclusion

When melt-spun Al-10.2 at.% Sm glass is heated, it first devitrifies into a cubic $\varepsilon$-$Al_{60}Sm_{11}$ phase. STEM images reveal profound spatial correlations of the Sm occupancy on the 16*f* Wyckoff positions in the non-stoichiometric $\varepsilon$ phase. Such spatial correlations cannot be reproduced by random occupation of these lattice sites, as suggested by Rietveld analysis of the X-ray diffraction spectrum based on. We perform both MC and MD simulations to try to interpret such spatial correlations. In MC simulations, a CE model is constructed to compute the free energy as a function of the Sm configuration on the partially occupied sites. The free energy predicted by the CE model is validated to be accurate within thermal fluctuations. The MC simulations successfully generate the thermodynamically favorable configurations, which do not reproduce the spatial correlations observed in experiment. With MD, we simulate the growth of the $\varepsilon$ phase by combining the liquid together with a crystal seed at an undercooled temperature. In addition to the conventional MD, we also run constrained MD in which the diffusion of Sm atoms is limited within a certain range $r_c$ from its initial position by an external potential. Our results show that when $r_c \sim 4$ Å, the as-grown phase shows spatial correlations that matches well with experiments, implying that the limited diffusivity of Sm is crucial for the appearance of the correlations of Sm content on the partially occupied sites.

## 6. Acknowledgement

This work was supported by the U.S. Department of Energy (DOE), Office of Science, Basic Energy Sciences, Materials Science and Engineering Division including a grant of computer time at the National Energy Research Supercomputing Center (NERSC) in Berkeley. Ames Laboratory is operated for the U.S. DOE by Iowa State University under contract # DE-AC02-07CH11358. The GPU-accelerated MD calculations were supported by the Laboratory Directed Research and Development (LDRD) program of Ames Laboratory.